\pdfoutput=1
\documentclass[11pt]{article}
\usepackage[margin=1in]{geometry}
\usepackage{amsmath,amssymb}
\usepackage{booktabs}
\usepackage{graphicx}
\usepackage{tikz}
\usepackage{sidecap}
\usetikzlibrary{arrows.meta,patterns}
\usepackage{microtype}
\usepackage{xspace}
\usepackage{url}
\usepackage[hidelinks]{hyperref}

\newcommand{\KSP}{\texttt{KSPSolve}\xspace}
\newcommand{\PtAP}{\texttt{PtAP}\xspace}

\newcommand{\MATBAIJ}{\texttt{MATBAIJKOKKOS}\xspace}
\newcommand{\nnzb}{\mathop{\mathrm{nnzb}}\nolimits}

\title{Reducing Data Movement in the Galerkin Product\\
of Block Algebraic Multigrid on GPUs}

\author{Mark F.\ Adams\\
Lawrence Berkeley National Laboratory\\
Berkeley, CA 94720\\
\texttt{mfadams@lbl.gov}}
\date{}

\begin{document}
\maketitle

\begin{abstract}
The Galerkin triple product $A_c = P^T A P$ dominates the recurring per-solve setup cost of
algebraic multigrid (AMG). For systems of PDEs the product is a \emph{rectangular-block}
sparse triple product --- $3\times3$ fine operators, $3\times6$ prolongators, $6\times6$
coarse operators for 3D elasticity --- a shape no vendor sparse library serves. We map the
algorithm space of this operator, from the classical two-pass product through fused,
schedule-reordered, and shared-memory-tiled variants to an inspector--executor form, under
an explicit DRAM/L2 traffic model, and implement the leading variants in portable Kokkos and
native CUDA backends using new PETSc\footnote{The public archive for this paper ---
source, extended version, measurement data, and reproduction configurations:
\url{https://gitlab.com/markadams4/petsc-claude-papers}.} blocked
matrix types. The model predicts per level which variant moves the fewest bytes; guided by
it, a tiled kernel with a sorted, search-free schedule and a column-split executor runs the
numeric product $1.8\times$ faster than the byte-minimal tile at byte-identical results, and
its portable Kokkos twin is within $19\,\%$ of the native kernel. We further present
prolongator filtering, a new GAMG algorithm that drops small blocks of the coarse-grid
space under a Frobenius criterion with a kernel-preserving projection. The driving
application is a fully GPU-resident blocked pipeline in PETSc --- finite-element assembly
through AMG setup and solve on primary blocked data, with no scalar expansion and no
operator-sized device--host transfers in the recurring phases. Weak-scaled to 64 GPUs at a
million unknowns per device, finite-element assembly and the solve run at 57--89\,\%
weak-scaling efficiency.
\end{abstract}


\section{Introduction}
\label{sec:intro}

Algebraic multigrid (AMG) solves the sparse linear systems of elliptic PDEs in near-linear
work and is a mainstay of extreme-scale simulation. Its cost structure splits into a
\emph{solve} phase dominated by sparse matrix--vector products (SpMV), a \emph{matrix
setup} phase dominated by the Galerkin triple product $A_c = P^T A P$ (hereafter \PtAP),
and a \emph{mesh setup} phase dominated by graph work and a matrix--matrix product in
smoothed aggregation~\cite{Vanek1996}. In nonlinear and time-dependent applications the
coarse-grid spaces are reused but the operator values change, so the \emph{numeric} \PtAP
recurs on every Newton or time step and its cost is paid as often as the solve
itself~\cite{Adams2010}.

For systems of PDEs the null space makes the prolongator rectangular-blocked: 3D elasticity
carries $3\times3$ fine blocks, $3\times6$ prolongator blocks, and $6\times6$ coarse
blocks. Vendor library support is at best partial: rocSPARSE~\cite{rocSPARSE} stores
rectangular blocks but provides no sparse--sparse product on them, cuSPARSE~\cite{cuSPARSE}
restricts block-sparse compute to square BSR, and Kokkos Kernels~\cite{Rajamanickam2021} is
square-block only. In previous work~\cite{AdamsBAIJ2026} we introduced a portable
Kokkos-backed blocked matrix type in PETSc~\cite{PETSc} (\MATBAIJ) and a conversion-free
blocked GAMG setup. Profiling that path shows the \PtAP numeric is L2-bandwidth-bound: its
blocked $A\cdot P$ stage runs at 89\,\% of A100 L2 bandwidth while DRAM sits at 20\,\% and
the floating-point pipes below 20\,\%, so the dense block contractions are cheap relative
to the bytes they move.

That observation frames this paper's question: \emph{across the space of algorithms for the
rectangular-block Galerkin product, which methods are most effective at reducing data
movement cost?} A companion question --- whether FP64 tensor cores can accelerate the block
contractions --- we answer analytically, in the negative: the largest contraction our block
shapes admit per fragment uses 42\,\% of an A100 FP64 tensor-core instruction, and $0.42$
of the $19.5$\,TF/s tensor-core peak falls below the $9.7$\,TF/s FMA peak~\cite{A100}
before staging costs. Reduced-precision paths retain a large advantage even at poor
fragment utilization; that accuracy-for-cost trade is future work.

\textbf{Contributions.}
\begin{enumerate}
\item A taxonomy and DRAM/L2 traffic model of the rectangular-block \PtAP algorithm space
  (two-pass, fused-recompute, scheduled, SMEM-tiled, inspector--executor), with per-level
  predictions and measured validation on A100 GPUs.
\item CUDA and Kokkos implementations of the leading variants using PETSc's new blocked
  matrix types --- atomic-free owner-computes contracts, tile-sorted schedules --- with a
  utilization audit of the byte-minimal kernel and the byte-identical column-split executor
  it motivates, $1.8\times$ faster on both backends.
\item Prolongator filtering in PETSc: a block-Frobenius filter with a kernel-preserving
  projection, evaluated as a trade of work and memory against convergence.
\item A fully GPU-resident blocked FEM-to-AMG pipeline in PETSc, from finite-element
  block-COO assembly through the solve~\cite{Adams2025}, with a weak-scaling study on
  1--64 GPUs organized around the four phases of a solve.
\end{enumerate}

An extended version of this paper carries the complete per-level data, the full
measurement narrative, and additional experiments; it is available in the author's paper
archive at \url{https://gitlab.com/markadams4/petsc-claude-papers}, as
\texttt{fem\_gamg/paper/extended.pdf}.
Appendices~\ref{app:repro} and~\ref{app:tables} hold the reproduction configuration and
secondary tables.

\section{Background and the Blocked Pipeline}
\label{sec:background}

\emph{Smoothed aggregation} (SA)~\cite{Vanek1996} aggregates strongly-coupled nodes,
builds a tentative prolongator $\tilde{P}$ that spans the user-provided near-null space
(the six rigid-body modes for 3D elasticity), smooths it with one damped-Jacobi step
$P = (I - \omega D^{-1} A)\tilde{P}$, and recurses on the Galerkin product
$A_c = P^T A P$. The SA coarse space carries one degree of freedom per null-space mode per
aggregate, so the block structure is not uniform across levels: the Galerkin numeric is a
chain of small dense contractions, $(6{\times}3)\cdot(3{\times}3)\cdot(3{\times}6)$ on the
finest level and $(6{\times}6)^3$ below it, threaded through sparse block-graph traversal.
Storing blocks natively amortizes one index over $bs_r bs_c$ values; the companion
paper~\cite{AdamsBAIJ2026} develops the blocked storage argument and the conversion-free
blocked GAMG setup this paper builds on.

\begin{figure}[t]
\centering
\includegraphics[width=0.7\linewidth]{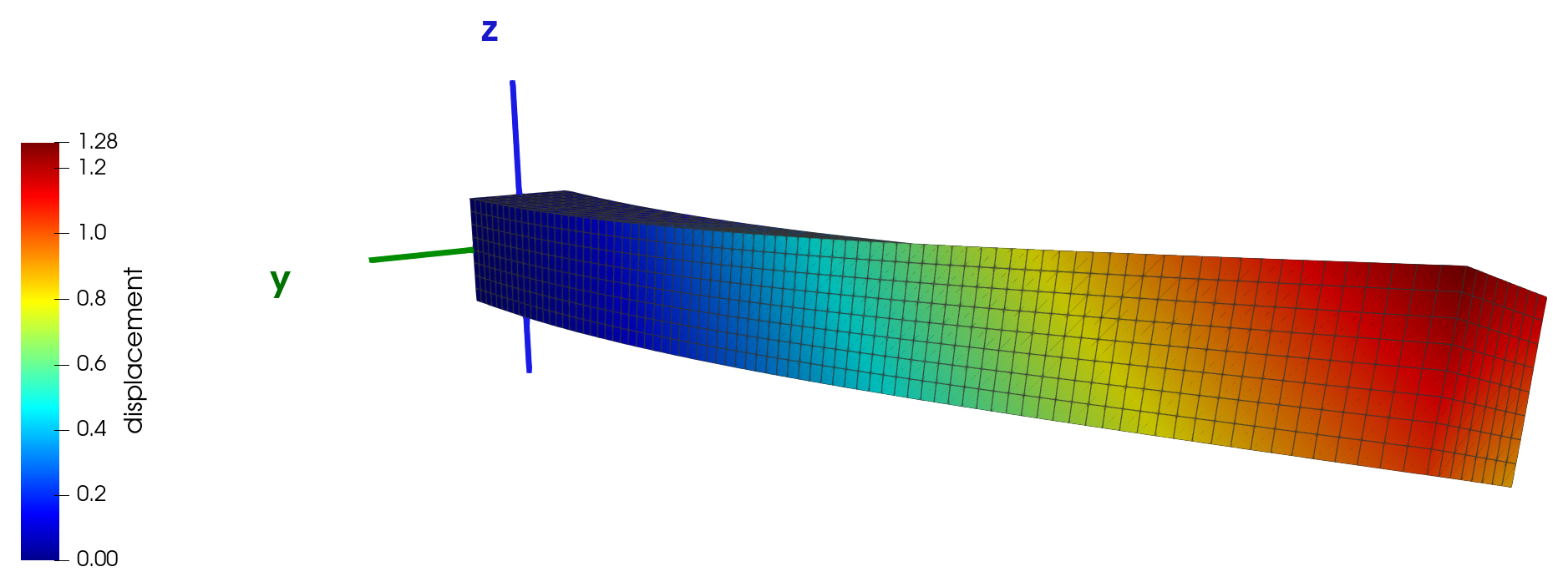}
\caption{Deformed cantilever beam under combined torsional and transverse end load,
colored by displacement magnitude (Q2 hexahedra, unit-cube elements; the torsional
load is increased here to make the twist visible).}
\label{fig:beam}
\end{figure}

\textbf{Test problem.} The driving application is 3D linear elasticity on a cantilever
beam (Figure~\ref{fig:beam}), Q2 hexahedral finite elements (PETSc
\texttt{snes/tutorials/ex56k}, \texttt{ex56cu}), clamped at one end with combined
torsional and transverse end load, solved with conjugate gradients preconditioned by AMG.
The beam is $10\times1\times1$ with $10k\times k\times k$ unit-cube elements. The
single-device base case fixes $k=16$: 1{,}045{,}440 displacement degrees of freedom, with
a frozen convergence gate (19 CG iterations, tip displacement 1.23166) that every variant
and backend must reproduce exactly, so a variant can only change time and memory, never
the mathematics.

\begin{figure}[t]
\centering
\resizebox{\textwidth}{!}{%
\begin{tikzpicture}[x=1cm, y=1cm,
  font=\footnotesize,
  box/.style={draw, rounded corners=2pt, align=center, inner sep=3pt,
              text width=29mm, minimum height=12mm},
  dev/.style={box, fill=blue!8, minimum height=17mm},
  hostb/.style={box, fill=orange!15, minimum height=9mm,
                postaction={pattern=north east lines, pattern color=orange!40}},
  lib/.style={box, fill=black!6, minimum height=10mm,
              postaction={pattern=crosshatch dots, pattern color=black!30}},
  libwide/.style={lib, text width=62mm},
  flow/.style={->, >={Stealth[length=2.5mm]}, thick},
  xfer/.style={<->, >={Stealth[length=2mm]}, dashed},
  rowlab/.style={font=\large\itshape, rotate=90, anchor=center}
]
\def\ca{0}\def\cb{3.8}\def\cc{7.6}\def\cd{11.4}\def\ce{15.2}
\node[dev] (d1) at (\ca,0) {\textbf{FEM assembly}\\ $A_e$ integration,\\ block-COO scatter};
\node[dev] (d2) at (\cb,0) {\textbf{AMG mesh setup}\\ graph, MIS aggregation,\\ $\tilde P$, smooth, filter:\\ emits $P_i$ per level};
\node[dev] (d3) at (\cc,0) {\textbf{matrix setup}\\ $A_c = P^{T}\!AP$ per level\\ (variants V0--V4)};
\node[dev] (d4) at (\cd,0) {\textbf{KSP solve}\\ CG on blocked\\ vectors};
\node[dev] (d5) at (\ce,0) {\textbf{AMG PC apply}\\ Chebyshev +\\ pb-Jacobi smoother,\\ $P$, $R$, coarse solve};
\node[hostb] (h1) at (\ca,2.3)  {mesh (\texttt{DMPlex});\\ FE tabulation $B$, $D$};
\node[hostb] (h2) at (\cb,2.3)  {residual\\ graph work};
\node[hostb] (h4) at (\cd,2.3)  {KSP/SNES/TS control\\ (iteration logic)};
\node[hostb] (h5) at (\ce,2.3)  {coarse LU\\ (option)};
\node[libwide] (l1) at ({(\ca+\cb)/2},-2.2)
  {native blocked device kernels\\ (\texttt{MATBAIJCUDA} / \MATBAIJ)};
\node[lib, text width=37mm] (l3) at (\cc+0.15,-2.2)
  {blocked SpGEMM (native);\\ $bs{=}1$: cuSPARSE, KK};
\node[libwide] (l4) at ({(\cd+\ce)/2},-2.2)
  {SpMV: native blocked, $bs{=}1 \to$ vendor CSR; coarse solve: cuDSS or host LU};
\draw[flow] (d1) -- (d2) node[midway, below=1.5mm]{$A_0$};
\draw[flow] (d2) -- (d3) node[midway, below=1.5mm, xshift=1.5mm, align=center]{$P_i$\\ $A_0$};
\draw[flow] (d3) -- (d4);
\draw[flow, <->] (d4) -- (d5);
\node[dev, dashed, text width=25mm, minimum height=7mm] (dS) at (\cc,1.68)
  {re-assemble $A$ (FEM stage)};
\draw[flow] (d4.north) to[bend right=15] node[below, font=\footnotesize]{$x_0$} (dS.east);
\draw[flow] (dS.south) -- node[left, font=\footnotesize, xshift=-0.5mm]{$A_0$} (d3.north);
\draw[xfer] (h1) -- (d1) node[midway, right, align=left]{maps, $B$, $D$\\ (one-time)};
\draw[xfer] (h2) -- (d2) node[midway, right]{one-time};
\draw[xfer] (h4) -- (d4) node[pos=0.22, right]{scalars};
\draw[xfer] (h5) -- (d5) node[midway, right, align=left]{one-time\\ factor};
\draw[dashed, thick, rounded corners=4pt, black!60]
  (5.5,-3.05) rectangle (17.15,3.1);
\node[anchor=north, font=\footnotesize\itshape, black!60] at (11.3,-3.12)
  {recurring Newton / time-step solve};
\node[rowlab] at (-1.95, 2.3)  {host};
\node[rowlab] at (-1.95, 0)    {device};
\node[rowlab] at (-1.95, -2.2) {libraries};
\end{tikzpicture}%
}
\caption{The blocked FEM--AMG workflow. The middle row is the device-resident pipeline,
the top row the host work, the bottom row the linear-algebra realization of each stage.
The host row is nearly empty by design: the dashed edges are the only host--device
transfers in the recurring phases and no operator-sized transfer crosses the bus. The
measured configuration factors the coarsest level on the device (cuDSS); the host coarse
LU is the fallback on non-NVIDIA hardware. The dashed enclosure marks the recurring
Newton/time-step solve, which re-assembles $A$ and refreshes the numeric products.}
\label{fig:workflow}
\end{figure}
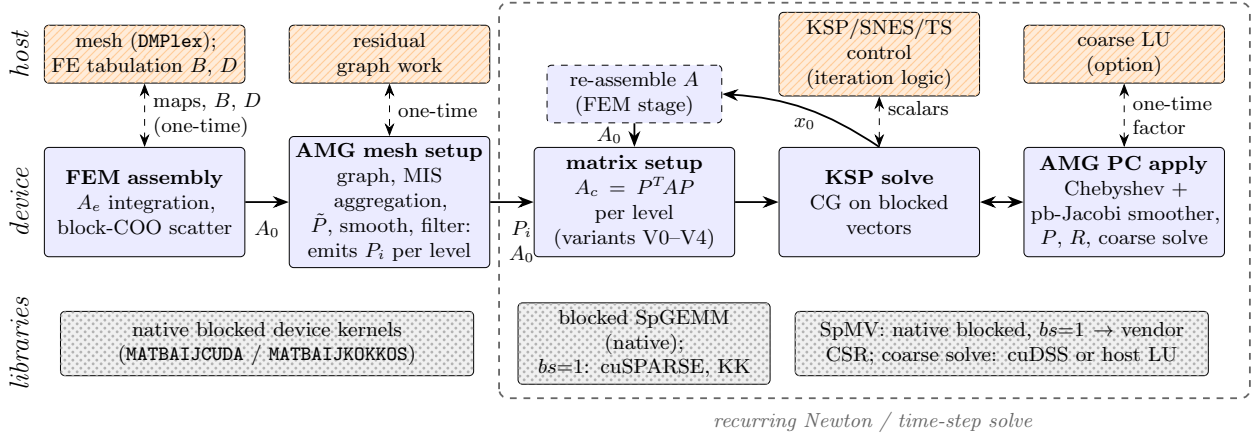

\textbf{The full-GPU blocked pipeline.} All measurements run inside a fully GPU-resident
blocked pipeline in PETSc (Figure~\ref{fig:workflow}) with two equally supported device
backends, one built on Kokkos (\MATBAIJ) and one directly on CUDA (\texttt{MATBAIJCUDA}),
selected at run time by a command-line option with no change to application code. A device
backend for \texttt{PetscFE} assembles the Jacobian and emits block COO directly into the
blocked matrix~\cite{AdamsBAIJ2026}; GAMG's aggregation (a device
maximal-independent-set coarsener), tentative prolongator, smoothing, filtering
(Section~\ref{sec:pfilter}), Galerkin products, and the preconditioned solve never leave
the device. Primary blocked data flows from element integration to the coarsest AMG level
without scalar expansion; the device--host transfer counters verify the recurring phases
run with zero copies in either direction (Section~\ref{sec:perf}).

\section{The Rectangular-Block \PtAP Algorithm Space}
\label{sec:space}

\subsection{Notation and traffic accounting}
Let $m_F$/$m_C$ be fine/coarse block rows; $Z_A$, $Z_P$, $Z_W$, $Z_C$ the block nonzero
counts of $A$, $P$, the intermediate $W = AP$, and $A_c$, with per-row averages $n_A$,
$n_P$, $n_C$; and $B_A, B_P, B_W, B_C$ the bytes of one dense block ($72$, $144$, $144$,
$288$ at fine-level shapes). Two derived quantities recur: $n_P$, the average number of
coarse aggregates each fine block row couples to, determines which algorithm is fastest;
and $w = Z_W/Z_P$, the fill growth of the intermediate, which we measure directly with a
counter in the symbolic phase (Table~\ref{tab:hier}). The model's conclusions are
insensitive to $w$.

Two counting rules generate every prediction. The \emph{stream rule}: a pass that touches
every block of an operand once costs its full byte size in DRAM, and an intermediate that
is written and read back costs double; summing streams gives each variant a \emph{DRAM
floor}. The \emph{touch rule}: in a sparse traversal every block touch is a memory request
served by whatever tier holds the block, and only misses become DRAM traffic; counting
touches gives a \emph{request volume}, and the serving tier determines which bandwidth
limits the kernel. The gap between floor and request volume quantifies the redundancy, and
the organizing question of the design space is where each algorithm puts its
redundancy. The A100's DRAM\,:\,L2\,:\,SMEM bandwidth ratio is roughly $1:3.2:12.5$
(Table~\ref{tab:machines}): the binding resource is L2 bandwidth, and moving redundant
requests from L2 into shared memory or registers gains another factor of ${\sim}4$.

\begin{table}[h!]
\centering
\caption{A100 machine parameters, from the whitepaper~\cite{A100} and microbenchmark
literature (L2).}
\label{tab:machines}
\small
\begin{tabular}{ll}
\toprule
 & NVIDIA A100-SXM4-40GB \\
\midrule
DRAM bandwidth & 1555\,GB/s \\
L2 cache & 40\,MB \\
L2 bandwidth & ${\sim}5$\,TB/s \\
Shared memory & 164\,KB/SM $\times$ 108 SMs \\
FP64 peak & 9.7\,TF/s FMA / 19.5\,TF/s tensor core \\
\bottomrule
\end{tabular}
\end{table}

\subsection{The variants}
\label{sec:variants}

\paragraph{V0: two-pass baseline.} Materialize $W = AP$ in DRAM, then $A_c = R\,W$ with
$R = P^T$ cached explicitly (its numeric refresh is a pure value gather, ${\sim}5\%$ of
the fine-level numeric). The floor is
$T_{V0} = B_A Z_A + 2 B_P Z_P + 2 B_W Z_W + B_C Z_C$; the write-and-read of $W$ is the
largest avoidable term. In practice V0 runs far from this floor: profiled, its $A\cdot P$
stage saturates L2 with gathered $P$-row requests (${\sim}n_P Z_A$ block touches) plus
hash work to place each contribution --- the binding resource is L2 requests and
discovery work, not DRAM streams or flops.

\paragraph{V1: fused recompute.} Never materialize $W$: the owner of coarse row $I$ walks
$R \to A \to P$ and accumulates directly into $A_c$~\cite{Bell2012}. The floor is lower
(no $W$ stream) and the ${\sim}1$\,GB $W$ allocation disappears, but the redundancy has a
precise coefficient: fine row $i$ is walked $n_P(i)$ times, so the flop count is exactly
$n_P\times$ V0's and the touch counts scale the same way. One aggregate's working set
(${\sim}0.3$\,MB) fits L2, so the redundant reads are served by L2 and V1 is limited by
the L2 request rate.

\paragraph{V2a/b/c: scheduling and tiling.} V2a reorders the fused traversal
(aggregate-contiguous or RCM coarse numbering) --- it can change which tier serves a
touch, never the touch count, a clean scheduling-vs-capacity experiment. V2b holds output
tiles in shared memory with a symbolic per-tile contribution list sorted by output slot,
eliminating per-hit search and all atomics; the tile is written once, coalesced. V2c
additionally stages each tile's $W$-panel on chip, reintroducing V0's intermediate
tile-locally: traffic approaches $(1+h)(B_A Z_A + B_P Z_P) + B_P Z_P + B_C Z_C$ where $h$,
a parameter of our model, is the halo fraction of fine rows shared by neighboring tiles
($1+h$ is the average number of tiles touching a fine row) --- the lowest floor of any
variant, if $h$ is small. On a regular grid $h$ is pure surface-to-volume: a tile of one
ideal $3^3$ aggregate touches its smoothed $5^3$ support, $h = 98/27 \approx 3.6$, and
$h = 0.25$ requires ${\sim}26^3$-point tiles, far beyond on-chip capacity.

\paragraph{V3: inspector--executor.} The symbolic phase flattens the numeric into one
tile-sorted stream of (operand, operand, output-slot) records, one per block multiply; the
numeric is a branch-free sweep with no hash, search, atomics, or traversal. The schedule
stream replaces the $W$ write-and-read at about half its bytes, and one inspection
amortizes over the many numeric refreshes of the production regime.

\paragraph{V4: transpose-free outer-product scatter.} Two applications of
$\mathrm{MTM}(A,B) = A^T B$ give $P^T A P$ without forming $R$. The device realization
reads $W'$ and $P$ once each, perfectly coalesced, and scatters block contributions into
$A_c$ by atomic accumulation: V0's streams with the amplification moved to the write side.
Its structural advantage is that parallelism ranges over fine rows in every pass,
uniform across levels --- which eliminates the coarse-level occupancy pathology measured
below.

\subsection{Model predictions}
\label{sec:model}

Table~\ref{tab:hier} gives the measured hierarchy of the base case and
Table~\ref{tab:model} evaluates the model on it. Three predictions follow. First, the
floors do not decide: the redundancy does. V1 holds the lowest floor on every level, yet
its request volume is $35\times$ its floor on the fine level and its flops are
$n_P\times$ V0's; equating the two costs gives a crossover at $n_P^{*} \approx
2.2$--$2.5$, and the measured $n_P = 4.8$--$8.0$ sits above the whole band, so V1 should
be slower by a factor of order $n_P$. (Unsmoothed aggregation, with $n_P = 1$, sits far below
the crossover, where V1 is the faster variant.) Second, scheduling alone has limited
value: if the redundant touches already hit L2, V2a can only convert DRAM refetches the
data does not have. Third, the lowest floor belongs to tiled fusion (V2c), governed by the
halo factor $h$; V4 matches V0's streams while parallelizing uniformly; V3 converts the
numeric into schedule-driven streaming for the reuse-dominated regime.

\begin{table}[h!]
\centering
\caption{Measured hierarchy of the 1.05M-DOF base case: three Galerkin products, block
units. $P$ is the smoothed, filtered (pf $=0.03$, Section~\ref{sec:pfilter}) prolongator;
$Z_W$ is the measured block count of $W = AP$ and $w = Z_W/Z_P$ its fill growth.}
\label{tab:hier}
\small
\setlength{\tabcolsep}{4.0pt}
\begin{tabular}{lccrrrrrrrrrr}
\toprule
product & $A$/$P$ blocks & & $m_F$ & $Z_A$ & $n_A$ & $Z_P$ & $n_P$ & $Z_W$ & $w$ & $m_C$ & $Z_C$ & $n_C$ \\
\midrule
l0 (fine) & $3{\times}3$ / $3{\times}6$ & & 348\,480 & 21.23M & 60.9 & 1.66M & \textbf{4.76} & 7.27M & 4.38 & 7\,243 & 627K & 86.6 \\
l1 & $6{\times}6$ / $6{\times}6$ & & 7\,243 & 627K & 86.6 & 51.6K & \textbf{7.12} & 376K & 7.29 & 1\,125 & 127K & 112.9 \\
l2 & $6{\times}6$ / $6{\times}6$ & & 1\,125 & 127K & 112.9 & 8\,970 & \textbf{7.97} & 29.4K & 3.27 & 98 & 3\,770 & 38.5 \\
\bottomrule
\end{tabular}
\end{table}

\begin{table}[h!]
\centering
\caption{Model-predicted traffic and footprints (GB) per Galerkin product for the
hierarchy of Table~\ref{tab:hier} at the measured $w$. ``Requests'' count block touches
(read side for V1, atomic read-modify-write side for V4). The V2c floor row is tabulated
at $h = 0.25$, the large-tile limit; on-chip tiles sit at single-aggregate granularity,
where the grid geometry gives $h \approx 3.6$ and the implemented variant measures exactly
that (Section~\ref{sec:perf-rap}).}
\label{tab:model}
\small
\begin{tabular}{lrrr}
\toprule
 & l0 & l1 & l2 \\
\midrule
V0 stream (floor) & 4.28 & 0.46 & 0.060 \\
V1 floor (no $W$) & 2.19 & 0.25 & 0.043 \\
V2c floor & 2.63 & 0.30 & 0.053 \\
V4 floor & 4.28 & 0.46 & 0.060 \\
V1 read requests & 76.6 & 10.4 & 2.6 \\
V4 write requests & 19.9 & 1.54 & 0.14 \\
\midrule
$W$ footprint & 1.05 & 0.108 & 0.008 \\
V3 schedule footprint (8 B/entry) & 1.09 & 0.057 & 0.010 \\
\midrule
flops V0 / V1 (GF) & 18.4 / 87.5 & 3.1 / 22.0 & 0.5 / 4.3 \\
\bottomrule
\end{tabular}
\end{table}

\subsection{Model vs.\ measured}
\label{sec:measured}

We validate the ordering predictions with V0 and V1 on the base case (one A100, 19
iterations everywhere), recording GPU-fenced \texttt{MatPtAPNumeric} times and per-kernel
DRAM/L2 bytes with Nsight Compute; a host control (an Apple M3, per-product stage timers)
supplies variant ratios free of GPU launch and occupancy effects. Tables~\ref{tab:wall}
and~\ref{tab:bytes} hold the data; four conclusions follow.

\begin{table}[h!]
\centering
\caption{Measured A100 wall time per Galerkin product, V0 vs.\ V1, with the per-product
$n_P$ of Table~\ref{tab:hier}. Per-product times are Nsight Compute kernel durations;
totals are GPU-fenced end-to-end \texttt{MatPtAPNumeric}.}
\label{tab:wall}
\small
\begin{tabular}{lrrrr}
\toprule
product & $n_P$ & V0 (ms) & V1 (ms) & ratio \\
\midrule
l0 & 4.76 & 82 & 570 & 7.0 \\
l1 & 7.12 & 12.3 & 106 & 8.6 \\
l2 & 7.97 & 4.1 & 139 & 33.9 \\
\midrule
total & & 87 & 684 & 7.9 \\
\bottomrule
\end{tabular}
\end{table}

\begin{table}[h!]
\centering
\caption{A100 measured bytes per Galerkin product (Nsight Compute, all kernels in the
numeric product), against the floors and request volumes of Table~\ref{tab:model}.}
\label{tab:bytes}
\small
\begin{tabular}{lrrrrr}
\toprule
 & \multicolumn{2}{c}{V0 (GB)} & \multicolumn{3}{c}{V1 (GB)} \\
\cmidrule(lr){2-3}\cmidrule(lr){4-6}
product & DRAM & L2 & DRAM & L2 & L2 hit \\
\midrule
l0 & 20.3 & 199 & 184.7 & 1246 & 89.1\,\% \\
l1 & 1.9 & 16.7 & 5.4 & 78 & 95.1\,\% \\
l2 & 0.2 & 2.8 & 0.11 & 16.6 & 99.4\,\% \\
\bottomrule
\end{tabular}
\end{table}

First, \textbf{the fused variant is slower by a factor of order $n_P$}, as predicted: $7.9\times$
end-to-end (Table~\ref{tab:wall}), with the host control reproducing per-product ratios
that track $n_P$ level by level --- the same coefficient on both memory systems, so the
verdict is a property of the algorithm, not the GPU. The variant with the lowest floor
moves the most DRAM bytes (Table~\ref{tab:bytes}), because request volume, not floor,
governs.

Second, \textbf{the model is reliable for ratios, orderings, and crossovers, not absolute
byte counts}. The measured L2 hit rate rises exactly as the per-level working set shrinks
into the 40\,MB L2 (89 to 99\,\%, Table~\ref{tab:bytes}), and the implied miss stream
reproduces 70--90\,\% of the measured DRAM traffic on every level; on the absolute scale,
measured L2 bytes exceed the block-touch count by roughly an order of magnitude (sector
granularity, index traffic, and accumulation read--modify--writes the touch count
excludes).

Third, \textbf{scheduling alone is insufficient}. RCM reordering of the fused kernel's
coarse rows moves fine-level DRAM by $-5.6\,\%$ and is wall-neutral; the natural
aggregate ordering is already good on this mesh, so by the model's dichotomy the
redundancy must be staged away (V2b/c), not reordered.

Fourth, \textbf{coarse levels fail on the GPU for occupancy, not traffic, reasons}: the
level-2 product is slower than the much larger level-1 product for V1 at a 99.4\,\% L2
hit rate, because one team per coarse row leaves most of the machine idle. The host
control shows normal scaling on the same products. Any variant therefore needs a
different launch geometry on coarse levels --- V4's fine-row parallelism delivers exactly
that by construction, and the tiled kernels' slot-based grids do the same. The MPI
analogue of this trade is GAMG's processor consolidation
(\texttt{-pc\_gamg\_process\_eq\_limit}, 1000 here), which gathers coarse operators onto
fewer processes as levels shrink: the per-process work there is so small that
deliberately idling processors loses little compute while cutting the message count and
the latency-dominated collectives --- on GPUs, where these coarse-level measurements show
a starved device already, that consolidation is critical.

When \texttt{-mat\_product\_algorithm} is left at its default the implementation selects
the variant per level from the level's own shape, following these measurements; naming a
variant forces it globally, which is how the comparisons in this paper are produced.

\section{Prolongator Filtering}
\label{sec:pfilter}

Smoothing the tentative prolongator densifies it, and that fill is squared through the
Galerkin product. We introduce \emph{prolongator filtering} in PETSc's GAMG
(\texttt{-pc\_gamg\_prolongator\_filter}): after smoothing, blocks of $P$ whose Frobenius
norm falls below a threshold are dropped, and a \emph{kernel-preserving projection}
restores the null space. Dropping entries with compensating lumping is established
practice~\cite{Vanek1996,Tuminaro2000}, but lumping preserves only the action on constant
vectors --- the translational modes; the projection here preserves the entire null space,
including the rotational modes, and to our knowledge this local post-filtering projection
is new. (Energy-minimizing prolongators~\cite{Olson2011} preserve the full null space
globally, by constrained minimization rather than local correction.)

Filtering perturbs the smoothed prolongator, so the correct accounting is possible
convergence degradation against reduced work and memory. Measured, the trade is strongly
favorable (Table~\ref{tab:pfsweep}): the iteration count is 19 at every threshold,
including unfiltered, while the hot \PtAP time falls $2.9\times$ at the paper's
$\text{pf}=0.03$ ($3.7\times$ at $0.05$, a value we have seen to be too high for some
problems), and $\nnzb(P)$, $\nnzb(W)$, and the coarse-operator fill --- every term of the
traffic model --- shrink with it.

\begin{table}[h!]
\centering
\caption{Prolongator-filter threshold sweep on the frozen base case (single A100, blocked
Kokkos backend; hot regime, three Galerkin products per solve). Newton = hot \KSP + hot
\texttt{MatPtAPNumeric}; memory is the 95th-percentile sampled device footprint. The
shipped defaults reproduce the band at faster absolute times (26.2\,ms per solve at
pf $=0.03$).}
\label{tab:pfsweep}
\small
\begin{tabular}{lrrrrr}
\toprule
threshold (pf) & iterations & \texttt{MatPtAPNumeric} hot (s) & Newton (s) & fine-$P$ nnz & p95 mem (GB) \\
\midrule
0 (off) & 19 & 0.230 & 0.706 & 47.4M & 10.2 \\
0.01    & 19 & 0.112 & 0.532 & 39.5M & 9.7 \\
0.02    & 19 & 0.091 & 0.501 & 33.8M & 9.5 \\
0.03    & 19 & 0.079 & 0.486 & 29.9M & 9.4 \\
0.05    & 19 & 0.061 & 0.448 & 24.4M & 9.1 \\
\bottomrule
\end{tabular}
\end{table}

\section{Implementation}
\label{sec:impl}

\subsection{CUDA: sorted schedules, tiles, and the column-split executor}
\label{sec:impl-cuda}
This section walks a series of optimizations --- owner-computes row kernels (V0/V1), the
tiled V2b, and the column-split executor that refines it --- each step keeping the
accumulation order, and hence the output bytes, of the one before; Section~\ref{sec:perf-rap}
measures the series as a ladder. The CUDA blocked types assign each output block row to
one owner, making the two-pass and
fused variants atomic-free with register-resident accumulation; the default team forms
split the owned row across a sub-warp and register-stage the operand blocks. The tiled
kernel V2b (the SMEM-tiled variant of Section~\ref{sec:variants}'s taxonomy) replaces row
ownership with \emph{output-slot} ownership: a symbolic pass
buckets every block contribution by its output slot in V0's iteration order, and the
numeric kernel owns a tile of slots, each thread accumulating one slot's contribution run
in registers with no search and no atomics, staging finished blocks through shared memory
for one coalesced writeback. Because every scalar element still sums its contributions in
exactly the two-pass order, V2b reproduces V0's accumulation bitwise on the host reference
(on device, FMA contraction differs between separately compiled kernels, so agreement is
to round-off). Wide coarse rows simply span several tiles; the one fallback is to V0 when a
single output block exceeds the register accumulator ($6\times6$).

The \emph{column-split} executor refines this once more: each slot is assigned to a group
of $bs_c$ lanes, one output column per lane, with slots assigned through a permutation
sorted by descending run length within each thread block's range. The serial accumulation
sequence per element is unchanged --- only the owning lane changes, verified bitwise. The
motivation and measured effect are in Section~\ref{sec:perf-rap}.

\subsection{MPI orchestration}
\label{sec:impl-mpi}
The parallel product gathers off-process rows of $P$ with a block-granular broadcast over
a dedicated \texttt{PetscSF}, runs the selected on-rank kernel on a merged local operator,
and assembles the result through the blocked COO path~\cite{AdamsBAIJ2026}; all maps and
plans are cached for values-only reuse. No stage of the distributed setup converts the
blocked operands to a scalar type. Communication-avoiding formulations are
analyzed in~\cite{Ballard2016}; our focus is the on-rank kernel.

\subsection{Kokkos backend and performance portability}
\label{sec:impl-kokkos}
The same algorithms express in Kokkos~\cite{Trott2022}, and the two backends form a
controlled portability comparison. Two lessons emerged. First, device performance is a
property of the control the backend exposes: the search-free range-parallel executor that
prevails on the host over-fetches $2.7\times$ on the device (a range policy gives each
output block one thread, so operand blocks cannot be kept resident across threads), and
the single-owner tile's Kokkos expression reaches the CUDA kernel's low traffic but not
its time --- the 48\,KB scratch tile admits one team per SM and occupancy collapses.
Second, the column-split executor expresses portably, because its resource
demands fit the constructs Kokkos exposes: no scratch tile, a 40-register accumulator, and
a lane-per-column layout that maps onto a \texttt{ThreadVectorRange} (power-of-two lane
padding is the one concession). On the base-case numeric product the Kokkos column-split
tile is within $19\,\%$ of the native-CUDA executor and both are their backends'
defaults, selected per product --- and therefore per level --- when
\texttt{-mat\_product\_algorithm} is unset.

\section{Performance}
\label{sec:perf}

All measurements are from NVIDIA A100-SXM4-40GB GPUs of the Perlmutter system at NERSC
(four per node; single-GPU runs on a dedicated node, multi-GPU runs on up to 16 nodes),
on the beam family of Section~\ref{sec:background} with the solver configuration held
fixed across every variant and backend (CG/AMG, Chebyshev point-block-Jacobi smoothing,
device cuDSS coarse solve; the full command and the cuDSS caveats are in
Appendix~\ref{app:repro}). Every configuration reproduces its scale's convergence gate
exactly. All A/B ratios are measured within one job with both legs on the same coarse
solver.

\subsection{The four phases of a solve, weak-scaled}
\label{sec:perf-phases}

The pipeline's cost divides into four parts: \emph{FEM matrix assembly} (recurs per
Newton step), \emph{mesh setup} (graph, aggregation, prolongators, symbolic products,
coarse factorization --- once per hierarchy), \emph{matrix setup} (the numeric Galerkin
products --- recurs), and the \emph{solve} (recurs). Table~\ref{tab:phases} weak-scales
all four at ${\sim}1$M unknowns per GPU, on both backends. This is the paper's top-level
result: the recurring phases weak-scale with no host excursions --- finite-element
assembly and the solve at 57--89\,\% efficiency (66--91\,\% on the Kokkos backend), the
numeric product setup at 68\,\% per product (the level count grows from 4 to 6, so its
per-step time scales at 41\,\%) --- and the
one-time mesh setup rises once at the transition from one GPU to MPI, where the
distributed graph and communication plans are first built, and then grows only $5\,\%$
from 8 to 64 GPUs (CUDA). The two heavy recurring phases are detailed in Sections~\ref{sec:perf-rap}
and~\ref{sec:perf-spmv}; FEM assembly we report only at this level, noting that its CUDA
and Kokkos kernels are near-identical twins at kernel granularity
(Table~\ref{tab:fepipe}, Appendix~\ref{app:tables}) --- the single-source pointwise
physics compiles to equivalent kernels through both programming models.

\begin{table}[h!]
\centering
\caption{The four phases of a solve, weak-scaled on the beam ($k = 16, 32, 64$ on 1, 8,
64 GPUs; 1.05M, 8.11M, 63.9M DOF; both device backends, Kokkos on its CUDA backend).
Recurring phases are hot-stage times (hierarchy warm, one Newton-step equivalent); mesh
setup is the cold setup of the same run minus its recurring part. Iterations (19, 19, 22)
and tip displacements are identical across backends and kernel variants at each scale, so
the solve row's efficiency is reported per iteration in the text. The $n=1$ legs are
dedicated-node jobs (the Kokkos leg re-measured after the tile-schedule build moved onto
the device); the MPI legs are one job per backend, whose same-job CUDA repeats agree with
the CUDA column within 1--3\,\%.}
\label{tab:phases}
\small
\begin{tabular}{lcrrrrrr}
\toprule
 & & \multicolumn{3}{c}{CUDA (GPUs)} & \multicolumn{3}{c}{Kokkos (CUDA) (GPUs)} \\
\cmidrule(lr){3-5}\cmidrule(lr){6-8}
phase (s) & recurs & 1 & 8 & 64 & 1 & 8 & 64 \\
\midrule
FEM matrix assembly   & per step & 0.334 & 0.363 & 0.377 & 0.359 & 0.380 & 0.395 \\
mesh setup            & once     & 1.44  & 5.62  & 5.93  & 1.10  & 2.82  & 3.43  \\
matrix setup (\PtAP)  & per step & 0.067 & 0.146 & 0.163 & 0.082 & 0.223 & 0.272 \\
solve (\KSP)          & per step & 0.209 & 0.298 & 0.427 & 0.349 & 0.458 & 0.615 \\
\bottomrule
\end{tabular}
\end{table}

\textbf{FEM assembly} weak-scales at 89\,\% efficiency to 64 GPUs. \textbf{Mesh setup} is
one-time; two changes removed its host excursions:
graph construction now runs as a device block collapse of the blocked operator ---
formerly the one place in the pipeline that still expanded it through a scalar
\texttt{AIJ} copy on the host --- and the coarse factorization runs on the device through
cuDSS. \textbf{Matrix setup}, the recurring numeric products, is
detailed in Section~\ref{sec:perf-rap}. \textbf{The solve} weak-scales per iteration at
70\,\% efficiency at 8 GPUs and 57\,\% at 64 (the single-GPU time per iteration over the
time per iteration at scale); the iteration count itself grows from 19 to 22 at 64 GPUs, identically on
both backends and under every kernel variant, so the growth is a property of the
coarsening at scale, not of the kernels. The fine level --- the blocked kernels --- is
62\,\% of the multigrid apply at 8 GPUs and 51\,\% at 64, and grows only 13\,\% per
iteration between them; the coarse solve is 6--7\,\% at both scales (the device
factorization is what makes it so: with the default host coarse LU, the coarsest
level alone was 51\,\% of the hot solve in a four-GPU probe, and cuDSS reduced that level
$19\times$ and the solve $1.90\times$). The remaining loss concentrates in the
middle levels, through two compounding effects. The hierarchy densifies with scale
--- operator complexity (total stored entries across levels over the fine level's) grows
1.143, 1.172, 1.189 --- so there is more coarse-level work per iteration; and the
partially consolidated middle levels carry severe load imbalance from GAMG's rank
consolidation --- max-over-ranks time
ratios of 13--15 there, against 1.1--1.2 on the fine level:
most ranks hold no rows there, and the full communicator waits. Retuning the coarsening
does not recover the iterations at acceptable cost (the parameter sweep is in
Appendix~\ref{app:tables}); the consolidated middle levels are the next algorithmic
target --- GAMG can already repartition the coarse grids, and evaluating that remedy,
repartitioning cost against the imbalance it removes, is future work.

The Kokkos column tracks the CUDA column's shape with two systematic differences. Its
mesh setup is cheaper at MPI scale (2.8 against 5.6\,s at 8 GPUs): the Kokkos
graph path is fully device-resident (\texttt{MatCreateGraph} is 0.17\,s at both MPI
scales, against 0.69--0.73\,s plus a host threshold-filter tail on CUDA). Its single-GPU
mesh setup halved (2.1 to 1.1\,s) when the tile-schedule build moved onto the device; the
MPI legs do not pay that build. Its recurring
phases are slower in absolute terms --- the merge-path SpMV is CUDA-only, so the Kokkos
solve runs the team apply kernel, and the native column-split executor leads its Kokkos
twin (Section~\ref{sec:impl-kokkos}) --- while scaling similarly: the Kokkos solve
weak-scales per iteration at
65\,\% efficiency at 64 GPUs against CUDA's 57\,\%, the slower baseline
leaving communication a smaller relative share.

\subsection{Matrix setup: the Galerkin numeric}
\label{sec:perf-rap}

The Kokkos and native-CUDA backends express the same algorithms, so comparing their
kernels on the fine-level product isolates kernel quality from algorithm choice.
Table~\ref{tab:kernelquality} reports the fine-level numeric product for the five
variants on both backends at matched launch geometry. The algorithm ordering is identical
across backends, as the model requires; the scheduled kernels (V2b tiled, V3 flat) are
the lowest-time products on CUDA; V1 is the negative result on both backends, its traffic
dominated by re-reading $A$ blocks across the prolongator fan-out. An attribution sweep
(operand staging on/off, sub-warp width, teams per block) shows operand staging and
launch geometry, not the backend, set the row-parallel kernels' traffic: staging cuts an
unstaged launch $8.5\times$ and matching geometry removes the remaining $2\times$, to
parity with the Kokkos team kernel on the $A\cdot P$ stage. Against the stage-consistent
streaming floor, the tiled kernel's $A\cdot P$ sits at $1.9\times$ and the full product
at $2.5\times$.

\begin{table}[h!]
\centering
\caption{Fine-level numeric \PtAP on both backends (A100, Nsight Compute, DRAM GB /
kernel ms; warp-cooperative CUDA kernels run at the Kokkos launch geometry so the
comparison isolates the backend). On Kokkos, V3 is the host-oriented range executor,
which over-fetches on the device (Section~\ref{sec:impl-kokkos}).}
\label{tab:kernelquality}
\small
\begin{tabular}{lrrrr}
\toprule
 & \multicolumn{2}{c}{Kokkos (team)} & \multicolumn{2}{c}{CUDA (warp-cooperative)} \\
\cmidrule(lr){2-3}\cmidrule(lr){4-5}
variant & DRAM (GB) & ms & DRAM (GB) & ms \\
\midrule
V0 (two-pass)      & 17.1  & 81.6  & 27.4  & 85.8  \\
V1 (fused)         & 184.3 & 573.8 & 407.8 & 678.0 \\
V4 (\texttt{mtm})  & 15.5  & 75.6  & 13.3  & 74.5  \\
V3 (flat)          & 46.2  & 217.0 & 10.5  & 48.1  \\
V2b (tiled)        & 13.3  & 168.7 & 10.5  & 45.4  \\
\bottomrule
\end{tabular}
\end{table}

\textbf{The column-split executor: the byte count is necessary but not sufficient.} The
byte-minimal tile saturates
no hardware resource: a per-counter audit locates the gap in the launch configuration the
single-owner design forces (a 96--166-register accumulator plus the 48\,KB tile admit one
or two thread blocks per SM, and unequal contribution-run lengths leave most lanes idle
--- $3$--$6\,\%$ of the machine's lanes doing work). The column-split executor
(Section~\ref{sec:impl-cuda}) removes all three limits at once --- 40 registers, no
scratch tile, run-length-balanced warps --- and Table~\ref{tab:colsplit} steps through the
ladder: $1.73\times$ on the numeric product at identical bytes, reproducing at
$1.79$--$1.83\times$ across the full hierarchy, and $1.47\times$/$1.43\times$ on the 8- and
64-GPU distributed product, whose on-rank stages run the same kernel. The window bound
matters: sorting within one thread block's slot range preserves schedule locality,
while a global sort costs more in broken reuse than balanced lanes recover.

\begin{table}[h!]
\centering
\caption{The column-split ladder on the base-case numeric \PtAP (A100; per-kernel columns
from Nsight Compute on the fine level; wall column is the device numeric per solve, all
levels). Each step is byte-identical to V0.}
\label{tab:colsplit}
\small
\begin{tabular}{lrrrrrr}
\toprule
executor & regs & occ.\,\% & lanes/32 & $A\,P$ ms & $R\,W$ ms & \PtAP ms/solve \\
\midrule
single-owner tile (V2b)      & 96  & 25 & 8.2  & 24.7 & 21.0 & 47.6 \\
\ + column split             & 44  & 46 & 12.4 & 19.0 & 11.7 & 33.8 \\
\ + windowed run-length sort & 44  & 48 & 23.5 & 14.0 & 9.8  & 28.3 \\
\ + operand prefetch         & 40  & 60 & 24.6 & 13.5 & 9.6  & 27.5 \\
\bottomrule
\end{tabular}
\end{table}

Two further variants close their questions by measurement. The inspector--executor V3
matches V2b within a few percent on the device hot numeric at a $1.7\times$ costlier
symbolic --- its compression advantage belongs to the host path, where the discovery work
it removes is expensive; on the device the simpler tiled schedule prevails. Implementing
V2c confirms the halo geometry: the best tile geometry measures $h = 3.6$, the
regular-grid value for single-aggregate tiles ($98/27$) --- equivalently, smoothed
aggregation's fan-out feeds each fine row to about five coarse rows, and no tile geometry
or reordering brings them into one tile; the fine-level traffic matches $(1+h)\times$
floor within $7\,\%$ --- the mechanism works exactly as designed and, at that $h$, the
panel staging costs more traffic than the $W$
stream it eliminates. V2c remains an explicit option, never selected by default.

\subsection{The solve: SpMV at the memory roof}
\label{sec:perf-spmv}

\begin{SCfigure}[0.95][t]
\centering
\includegraphics[width=0.6\linewidth]{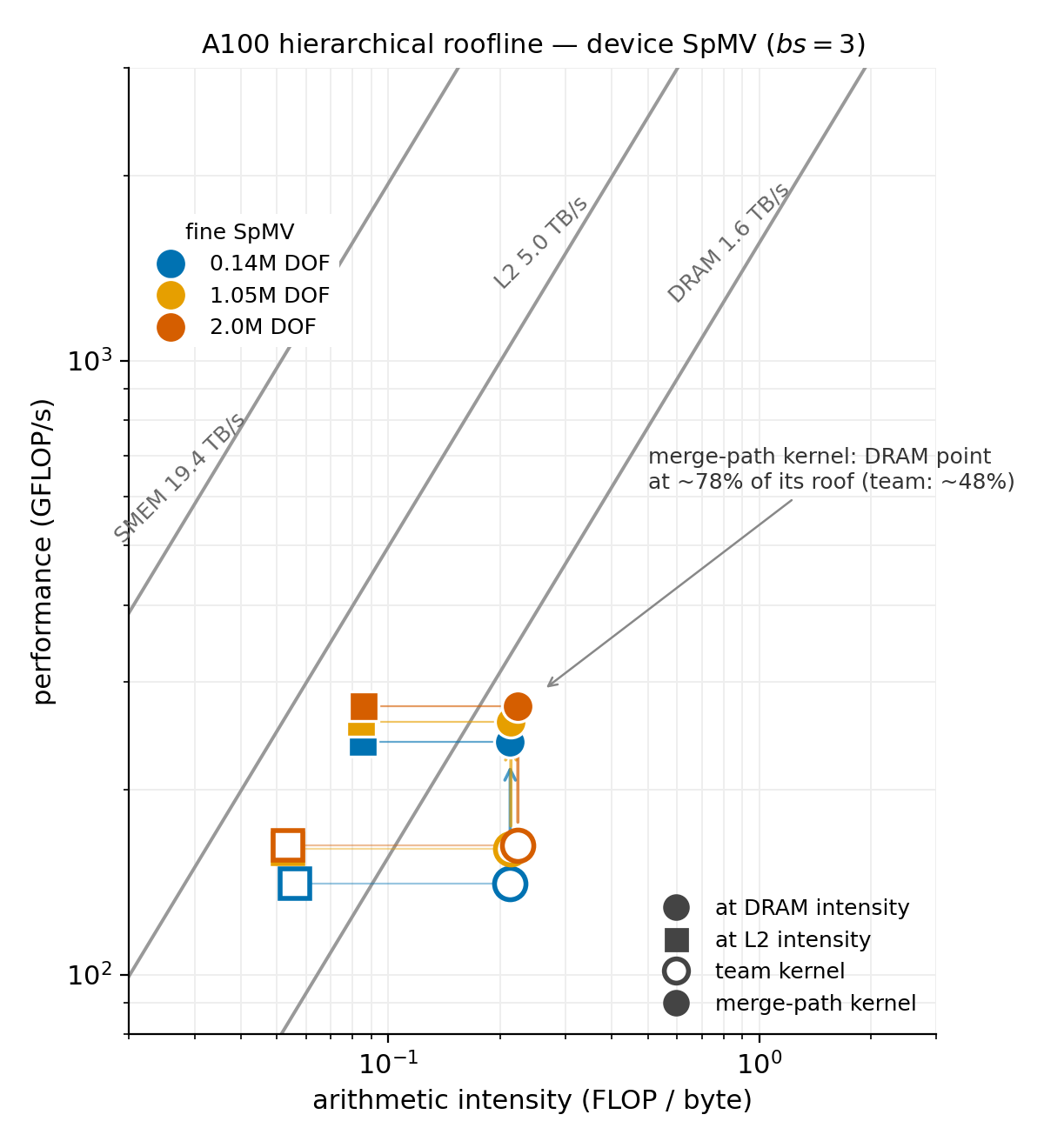}
\caption{A100 roofline of the blocked device SpMV ($bs{=}3$, \texttt{baijcuda} apply),
measured with Nsight Compute at three sizes ($0.14$--$2.0$\,M DOF); both kernels are
plotted at useful flops ($2\times$nnz). The team kernel (hollow) sits at ${\sim}48\,\%$
of the DRAM roof with L2 nearest its roof --- latency-bound, invariant in size. The
merge-path kernel (filled) moves the same operator, at the same arithmetic intensity, to
${\sim}78\,\%$ of the DRAM roof: the apply becomes bandwidth-bound.}
\label{fig:spmv-roofline}
\end{SCfigure}

The multigrid apply is dominated by the blocked SpMV in the Chebyshev smoother. Our
original kernel --- a half-warp team per block row, each lane walking its own block ---
is size-invariantly latency-bound at ${\sim}48\,\%$ of the DRAM roof with L2 the busiest
tier: with each lane reading its own 72\,B block, a load uses one 8\,B element per 32\,B
sector, and sector replay saturates the request path while DRAM idles. No kernel-level
adjustment (shuffle reductions, team widths, vectorized loads, operand staging) moved the
wall clock: within the row-parallel pattern the request stream itself is the limiter, and
reaching the roof --- which a well-implemented SpMV should --- required a new access pattern, not
further tuning. The merge-path kernel that replaces it partitions the block
stream into equal, block-aligned per-thread ranges: block positions
become compile-time constants, one load of $x$ serves $bs$ rows, values load with 16\,B
vector instructions, and a warp-level segmented reduction combines row partials before
touching memory, cutting the $y$-vector atomics ${\sim}18\times$.
Figure~\ref{fig:spmv-roofline} shows the fine-level operating point moving to the memory
roof: ${\sim}78\,\%$ of peak DRAM bandwidth, a $1.45$--$1.48\times$ aggregate
\texttt{MatMult} gain across the hierarchy at unchanged iteration counts, and
$1.22$--$1.42\times$ on the hot solve end to end at 0.14--2.0M DOF ($1.31\times$ and
$1.29\times$ on the 8- and 64-GPU beams). The apply is now bandwidth-bound in the
ordinary sense; the remaining headroom is the last quarter of the roof, and beyond it,
further gains require shrinking the operator stream itself (reduced-precision storage).
The merge-path kernel is the default
for square blocks $2 \le bs \le 6$.

\subsection{Weak scaling of the solve}
\label{sec:perf-weak}

Table~\ref{tab:weak} completes the solve row of Table~\ref{tab:phases} with the series
detail. The kernel A/B ratios persist at every scale (merge SpMV $1.29$--$1.39\times$ on the hot
solve, column split $1.43$--$1.79\times$ on the numeric product), so the scaling loss is
not a kernel property; the attribution of Section~\ref{sec:perf-phases} --- fine level
and blocked kernels scale, consolidated middle levels do not --- is the finding. The
device--host transfer counters verify residency at scale: no recurring phase moves an
operator across the bus, and the hot solve runs with zero copies in either direction.

\begin{table}[h!]
\centering
\caption{Weak scaling of the hot solve (hierarchy built and warm; A100s, four per node).
Iterations are backend- and variant-independent at each scale; the tip displacement
(1.23166, 1.23179, 1.23185) converges with the mesh. Efficiency is per iteration,
relative to one GPU.}
\label{tab:weak}
\small
\begin{tabular}{rrrrrrrr}
\toprule
GPUs & mesh & DOF & levels & its & hot \KSP (s) & ms/iteration & efficiency \\
\midrule
1  & $160{\times}16{\times}16$ & 1.05M & 4 & 19 & 0.209 & 11.0 & 100\,\% \\
8  & $320{\times}32{\times}32$ & 8.11M & 5 & 19 & 0.298 & 15.7 & 70\,\% \\
64 & $640{\times}64{\times}64$ & 63.9M & 6 & 22 & 0.427 & 19.4 & 57\,\% \\
\bottomrule
\end{tabular}
\end{table}

\section{Conclusion and Future Work}
\label{sec:conclusion}

This paper treated the rectangular-block Galerkin product as an algorithm space rather
than a single kernel. A traffic model separating a mandatory data floor from the request
volume a schedule actually issues predicts, per level, which variant has the lower cost
as a function of one geometric ratio, $n_P$; the predictions hold on the A100, with a
host control confirming the fused variant's loss coefficient is a property of the
algorithm, not the GPU. The model is reliable for ratios, orderings, and crossovers, not
absolutes. The byte count proved necessary but not sufficient: a utilization audit of the
byte-minimal tile led to the column-split executor --- byte-identical by construction ---
$1.8\times$ faster on both backends and within $19\,\%$ of native in its portable form,
now both backends' defaults. Prolongator filtering, with a block null-space-preserving
correction that retains the rotational modes a scalar lumping would drop, controls
operator complexity at unchanged iteration counts and reduces the device memory
footprint.
The merge-path restructuring of the blocked SpMV moved the apply from a latency floor at
half the DRAM roof to ${\sim}78\,\%$ of it. Weak-scaled to 64 GPUs at a
million unknowns per device, finite-element assembly and the solve weak-scale at
57--89\,\% efficiency with
no host excursions in any recurring phase; the loss concentrates in the densifying,
partially consolidated middle levels of the algebraic hierarchy, which names the next
algorithmic target.

Future work: reduced-precision Galerkin products and coarse levels (where the tensor
cores' reduced-precision paths retain a large advantage, and multigrid's algebraic error
structure is tolerant~\cite{McCormick2021}); a systematic sweep of the
process-consolidation control at scale, which the measured mid-level imbalance now
motivates directly; aggregate-aligned additive-Schwarz smoothers spanning process
boundaries; a dedicated anisotropy study; communication-avoiding formulations of the
blocked product~\cite{Ballard2016}; and H100-class re-tuning~\cite{A100}.

\section*{Code and Data Availability}

The implementation is developed in a fork of PETSc. The public archive for this paper is
\url{https://gitlab.com/markadams4/petsc-claude-papers}, and it holds: the source, as a
patch against a pinned upstream PETSc commit that reconstructs the measured tree exactly
(\texttt{code/}); the extended version, with the complete per-level data and additional
measurements (\texttt{fem\_gamg/\allowbreak paper/extended.pdf}); the measurement data behind every
figure and table, with the job identifiers and the reduction that produced each row
(\texttt{fem\_gamg/data/}); and the test problems, run configurations, and documentation
for reproducing the experiments (\texttt{fem\_gamg/REPRODUCE.md}; Appendix~\ref{app:repro}
holds the canonical command). A snapshot with an archival DOI will accompany the final
version.

\section*{Use of AI}

Anthropic's Claude models, operating agentically through Claude Code, were used under the
author's direction for background research and exploration, analysis, code writing,
debugging, and code optimization, and for drafting of the manuscript. In particular, the
data-movement model of Section~\ref{sec:space} and its measurement-based validation were
AI-formulated. The author made all design and experimental decisions, verified all
results, and takes full responsibility for the content.

\section*{Acknowledgments}

Thanks to the PETSc team for developing a well-engineered numerical library that provided an
ideal basis for the extensions developed in this project. This material is based upon work
supported by the U.S. Department of Energy, Office of Science, Office of Advanced Scientific
Computing Research, Scientific Discovery through Advanced Computing (SciDAC) Program through
the FASTMath Institute, under contract number DE-AC02-05CH11231 at Lawrence Berkeley National
Laboratory. This research used resources of the National Energy Research Scientific Computing
Center (NERSC), a U.S.\ Department of Energy Office of Science user facility located at
Lawrence Berkeley National Laboratory.

\appendix

\section{Reproduction Configuration}
\label{app:repro}

The canonical run command is below (PETSc \texttt{ex56cu}, native-CUDA backend). The
Kokkos twin substitutes \texttt{baijkokkos}/\texttt{kokkos} for the two type options; the
weak-scaling runs change only \texttt{-dm\_plex\_box\_faces} to $320{,}32{,}32$ and
$640{,}64{,}64$.
{\small
\begin{verbatim}
./ex56cu -dm_plex_dim 3 -dm_plex_shape zbox -dm_plex_simplex 0
  -dm_plex_box_faces 160,16,16 -dm_plex_box_lower 0,-0.5,-0.5
  -dm_plex_box_upper 10,0.5,0.5 -lx 10 -petscspace_degree 2 -run_type 4
  -dm_mat_type baijcuda -dm_vec_type cuda -snes_type ksponly -snes_max_it 1
  -n_solves 10 -ksp_type cg -ksp_norm_type unpreconditioned -ksp_rtol 1.e-8
  -pc_type gamg -pc_gamg_aggressive_coarsening 1 -pc_gamg_threshold 0.05
  -pc_gamg_threshold_scale 0.5 -pc_gamg_coarse_eq_limit 2000
  -pc_gamg_process_eq_limit 1000 -pc_gamg_prolongator_filter 0.03
  -pc_gamg_prolongator_filter_scale 0.5 -mg_levels_pc_type pbjacobi
  -mg_coarse_ksp_type preonly -mg_coarse_pc_type bjacobi
  -mg_coarse_sub_pc_factor_mat_solver_type cudss -log_view -log_view_gpu_time
\end{verbatim}
}
Ten solves are what separates the one-time mesh setup from the recurring per-step phases
of Section~\ref{sec:perf-phases}: the first solve is the cold one, and every recurring
number is the hot-stage average of those that follow, read from the \texttt{-log\_view}
Solve stage. \texttt{-mat\_product\_algorithm} names the product variant under study;
left unset, the implementation selects per level (Section~\ref{sec:impl-kokkos}). The coarse level is
factored on the device through cuDSS, an NVIDIA library (\texttt{--download-cudss}) and a
reproduction dependency: omitting the flag silently selects a host LU and changes the
solve-phase numbers. cuDSS is CUDA-only, so on non-NVIDIA hardware the Kokkos backend
falls back to a host LU factorization of the few-hundred-row coarsest operator. The
convergence gates are 19 iterations / tip displacement 1.23166 at $k=16$ ($n=1$), 19 /
1.23179 at $k=32$ ($n=8$), and 22 / 1.23185 at $k=64$ ($n=64$); every kernel variant and
both backends must reproduce them exactly.

\section{Additional Measurements}
\label{app:tables}

\paragraph{Shared-memory tile budget.} The single-owner tile width is set so a tile's
output blocks fit the static shared-memory budget (Table~\ref{tab:smem}); the
column-split executor needs no staging tile, which unties slots-per-block from this
budget and is part of its occupancy gain.

\begin{table}[h!]
\centering
\caption{Shared-memory tile budget (FP64): output blocks per tile at the default 48\,KB
static shared memory and the dynamic opt-ins, for the block shapes of the elasticity
hierarchy. The register accumulator bounds a single output block at $6\times6$.}
\label{tab:smem}
\small
\begin{tabular}{lrrrr}
\toprule
output block & B/block & cap @48\,KB & cap @64\,KB & cap @164\,KB \\
\midrule
$1\times1$ (scalar) & 8   & 6000 & 8000 & 20500 \\
$3\times3$          & 72  & 666  & 888  & 2277  \\
$3\times6$ ($W$)    & 144 & 333  & 444  & 1138  \\
$6\times6$ ($A_c$)  & 288 & 166  & 227  & 569   \\
\bottomrule
\end{tabular}
\end{table}

\paragraph{FEM assembly kernel twins.} Table~\ref{tab:fepipe} supports the portability
claim of Section~\ref{sec:perf-phases}: the two backends' assembly kernels are
near-identical at kernel granularity.

\begin{table}[h!]
\centering
\caption{Finite-element assembly and residual on the base case (A100, Nsight Compute,
DRAM GB / kernel ms).}
\label{tab:fepipe}
\small
\begin{tabular}{lrrrr}
\toprule
 & \multicolumn{2}{c}{CUDA} & \multicolumn{2}{c}{Kokkos} \\
\cmidrule(lr){2-3}\cmidrule(lr){4-5}
kernel & GB & ms & GB & ms \\
\midrule
Jacobian integration ($A_e$)   & 221.5 & 370.9 & 238.7 & 387.6 \\
blocked scatter $\to$ COO      & 12.7  & 24.1  & 12.9  & 24.3  \\
\texttt{MatSetValuesCOO} (fine)& 4.08  & 4.3   & 4.08  & 4.2   \\
residual (volume)              & 4.09  & --    & 4.14  & --    \\
\bottomrule
\end{tabular}
\end{table}

\paragraph{Coarsening parameters at 64 GPUs.} The 19-to-22 iteration growth invites
retuning;
Table~\ref{tab:knobs} sweeps the natural controls within one job. Three settings recover
two
iterations but cost far more than they save; one --- a second pass of aggressive
coarsening --- keeps 22 iterations and runs 14\,\% faster per solve, a shallower, cheaper
hierarchy at unchanged convergence. We report these as observations; the base
configuration is retained throughout the paper.

\begin{table}[h!]
\centering
\caption{Coarsening-parameter sweep at 64 GPUs (one job; hot \KSP per solve, ratio to the
job's base run; option names abbreviate the \texttt{-pc\_gamg\_} prefix). The iteration
growth is backend-independent (the Kokkos backend also converges in 22).}
\label{tab:knobs}
\small
\begin{tabular}{lrrr}
\toprule
configuration & its & hot \KSP (s) & vs.\ base \\
\midrule
base (Appendix~\ref{app:repro})           & 22 & 0.466 & 1.0 \\
\texttt{threshold 0.075}                  & 20 & 0.661 & $1.42\times$ slower \\
\texttt{threshold\_scale 1.0}             & 20 & 0.759 & $1.63\times$ slower \\
\texttt{coarse\_eq\_limit 5000}           & 20 & 1.288 & $2.76\times$ slower \\
\texttt{aggressive\_coarsening 2}         & 22 & 0.402 & $1.16\times$ faster \\
\texttt{process\_eq\_limit 4000}          & 22 & 0.451 & $1.03\times$ faster \\
\bottomrule
\end{tabular}
\end{table}

\bibliographystyle{plain}
\bibliography{refs}

\end{document}